# Pressure-induced superconductivity and structural transition in ferromagnetic $Cr_2Si_2Te_6$

Wanping Cai[1], Hualei Sun[1,*], Wei Xia[2], Changwei Wu[1], Ying Liu[3], Jia Yu[1], Junjie Yin[1], Hui Liu[1], Yu Gong[4], Dao-Xin Yao[1], Yanfeng Guo[2], and Meng Wang[1,#]

[1]*School of Physics, Sun Yat-Sen University, Guangzhou, Guangdong 510275, China*
[2]*School of Physical Science and Technology, ShanghaiTech University, Shanghai 201210, China*
[3]*Xi'an Mordern Chemistry Research Institute, Xi'an, 710065, China*
[4]*Beijing Synchrotron Radiation Facility, Institute of High Energy Physics, Chinese Academy of Sciences, Beijing 100049, China*

Abstract:

The discovery of intrinsic magnetism in atomically thin two-dimensional transition-metal trichalcogenides has attracted intense research interest due to the exotic properties of magnetism and potential applications in devices. Pressure has proven to be an effective tool to manipulate the crystal and electronic structures of the materials. Here, we report investigations on ferromagnetic van der Waals $Cr_2Si_2Te_6$ via high-pressure synchrotron x-ray diffraction, electrical resistance, Hall resistance, and magnetoresistance measurements. Under compression, $Cr_2Si_2Te_6$ simultaneously undergoes a structural transition, emergence of superconductivity at 3 K, sign change of the magnetoresistance, and dramatic change of the Hall coefficient at ~8 GPa. The superconductivity persists up to the highest measured pressure of 47.1 GPa with a maximum $T_c$ = 4.5 K at ~30 GPa. The discovery of superconductivity in the two-dimensional van der Waals ferromagnetic Cr-based $Cr_2Si_2Te_6$ provides new perspectives to explore superconductivity and the interplay between superconductivity and magnetism.

Main article:

Layered structural materials have been a fertile playground to investigate mechanisms of fundamental physics and explore potential applications[1–14]. The copper oxide and iron based high temperature superconductors (HT$_c$s) with strong electronic correlation and weak magnetic exchange interactions between layers attracted intensive research in the past decades because of the emergence of unconventional superconductivity (SC) and intimate interplay of magnetism and SC[15,16]. Another family of materials with two-dimensional (2D) layers linked by van der Waals (vdW) forces has received growing attention, owning to the diversity of the system, for example, nonmagnetic transition-metal dichalcogenides $MX_2$ ($M$ = Mo, W, and $X$ = S[17,18], Se[19], and Te[20,21]), antiferromagnetic (AFM) Fe$PY_3$ ($Y$ = S[22], Se[23]), ferromagnetic (FM) Cr$Z$Te$_3$ ($Z$ = Si[9], Ge[25]) and CrI$_3$[2,26]. This family also has extraordinary properties including thermoelectricity[28], large magnetoresistance (LMR)[21], intrinsic magnetic order even down to monolayer[2], etc. These exotic discoveries motivate further scientific and engineering explorations on the 2D vdW family.

Pressure is an ideal tool to tune crystal and electronic structures as well as magnetism without simultaneously introducing any disorders[29,30]. Previous high-pressure studies have found insulating-metal transition (IMT), structural transition, and spin-crossover in AFM FePY$_3$ (Y = S, Se), and emergence of SC in FePSe$_3$[22,23]. For nonmagnetic 2D vdW WTe$_2$ with LMR, SC has also been observed above ~10 GPa without a structural transition[20,21]. Regarding the search for SC, it is of great interest for FM materials because the FM compounds may host spin-triplet SC as observed in Sr$_2$RuO$_4$[31] and UTe$_2$[32–35].

In the pursuit of SC in FM materials, Cr$_2$Si$_2$Te$_6$ is an ideal member of the 2D vdW FM insulators[24]. It exhibits a 2D Ising FM order with moments of 2.73 $\mu_B$ per Cr$^{3+}$ and $T_N$ = 32 K lying ferromagnetically along the $c$ axis[36,37]. The Cr$^{3+}$ ions have the electronic configuration of $4s^03d^3$, and the octahedral environment splits the $d$ levels in the $t_{2g}$ and $e_g$ manifolds. According to the Hund's rule, the Cr$^{3+}$ is in the $S$ = 3/2 state with three polarized electrons in the 3 $t_{2g}$ orbitals[38,39]. We note that SC has been observed in CrAs under pressure[40] and $A_2$Cr$_3$As$_3$ ($A$ = alkali metals) at ambient pressure[41], where the Cr$^{3+}$ ions exhibit similar spin configuration.

In this work, Cr$_2$Si$_2$Te$_6$ was investigated via *in situ* high-pressure synchrotron x-ray diffraction (XRD) and electrical resistance in diamond anvil cells (DAC) with pressure up to 47.1 GPa. We found that a pressure-induced structure transition occurs at ~7 GPa, and SC appears concomitantly. The superconducting transition temperature $T_c$ expands from ~3 − 4.5 K in the pressure range from 8.0 to 47.1 GPa, where the highest pressure we achieved. At the pressure of the structural transition, there are a sign change from negative to positive of the magnetoresistance (MR) and an abrupt change of the Hall coefficient, indicating a reconstruction of the Fermi surfaces. A temperature-pressure ($T$-$P$) phase diagram is established.

**Results**
**Pressure-induced structural transition.** Cr$_2$Si$_2$Te$_6$ has a 2D honeycomb layered structure, which is formed by edges sharing CrTe$_3$ octahedra. Each Cr$^{3+}$ ion is octahedrally coordinated by six nearest neighbor Te atoms, as shown in Fig. 1a. The Cr$^{3+}$ ions are arranged in a nearly perfect honeycomb lattice staked along the $c$ axis. Figures 1b and 1c display measurements of XRD and magnetic susceptibility on single crystals of Cr$_2$Si$_2$Te$_6$. The results reveal high quality of our sample and a FM ordered temperature of $T_N$ = 32 K that is consistent with previous reports[7,9].

From the XRD patterns under pressure in Figs. 2a and 2b, it is obvious to see a phase transition from the discontinuous shifts of the diffraction peaks at $Q$ = (0 0 6), (1 1 3), and (1 1 6). The structure transition occurs at approximately 7 GPa. The low-pressure (LP) phase retains the ambient pressure crystal structure of $R$-3 up to 8 GPa, then evolves into a high-pressure (HP) phase. The XRD patterns with limited peaks under high pressure prevent us to refine the corresponding structure precisely. We notice that Cr$_2$Ge$_2$Te$_6$ undergoes a structural transition from $R$-3 to $R$3 under pressure[42]. The reflection pattern of Cr$_2$Si$_2$Te$_6$ at high pressure could be fitted by the $R$3 structure with preferred orientation. We adopted the $R$3 structure to fit the HP phase of Cr$_2$Si$_2$Te$_6$. The determined volume and lattice constants are plotted in Figs. 2c and 2d.

**Pressure-induced insulator-metal transition and superconductivity.** Figures 3a-3d show the typical temperature dependence of electrical resistance for single crystals of Cr$_2$Si$_2$Te$_6$ measured in a DAC in the pressure range of 1.0 − 47.1 GPa. At ambient pressure, Cr$_2$Si$_2$Te$_6$ is suggested to be a Mott insulator with an indirect gap of 0.4 eV and a direct gap of 1.2 eV. Upon applying a pressure of 1.0 GPa, a hump in resistance appears, yielding an IMT at $T_{hump}$ = 69 K (Fig. 3b). The transition temperature $T_{hump}$ evolves to the higher temperature upon increasing pressure. At pressures above 5 GPa, $T_{hump}$ moves out of the measured temperature range, demonstrating that $Cr_2Si_2Te_6$ becomes a metal. Interestingly, with further increasing pressure, an abrupt drop in resistance appears at ~8.0 GPa, demonstrating the emergence of SC.

The data in Fig. 3c show clear pressure-induced SC with zero resistance starting from 9.1 GPa and persisting up to 47.1 GPa which is the highest pressure we achieved in our measurement. The SC transition temperature $T_c$ is 3.3 K at 9.1 GPa, defined by the intersection of the tangent to the resistance curve during the transition process and the straight-line fit of the normal state above the transition. The $T_c$ reaches 4.0 K at 19.5 GPa. For higher pressures above 26.0 GPa, the SC transitions are broadened, and the transitions with zero resistance have a clear trend moving to lower temperatures. At 43.8 GPa, the resistance does not reach zero at 2 K. However, the temperature of the drop in resistance is close to the maximum $T_c$.

To determine the upper critical field $H_{c2}$, we present temperature dependence of the resistance at 19.5 GPa under different magnetic fields in Fig. 3d. The $T_c$s are clearly suppressed by magnetic field. Zero resistance could not be observed at 3 and 4 T because the lowest temperature we could reach in our experiment is 2 K. The Ginzburg-Landau formula is adopted to fit the experimental data, yielding $H_{c2}$ = 4.0 T (Fig. 3e).

**An elaborate temperature-pressure phase diagram**. We summarize the data extracted from the pressure dependent XRD and resistance in a temperature-pressure (*T-P*) phase diagram as presented in Fig. 3f. Before the structural transition, the insulating FM ground state undergoes an IMT. Our detailed measurements suggest that the FM order and ordered temperature $T_N$ persist closely to the structural transition. The SC appears simultaneously as the structural transition, where the FM order may not survive (Supplementary figures).

**Pressure-induced sign change of magnetoresistance**. Previous studies have reported that $Cr_2Si_2Te_6$ exhibits a negative magnetoresistance (MR) at ambient pressure[24]. The MR is related to the topology of Fermi surfaces and would have connections to SC. To study the pressure dependence of the MR, we measured magnetic field dependence of the resistance under fixed pressures at 10 K, as shown in Fig. 4a. The results reveal a dramatic change of the MR below 9.1 GPa and a negative to positive sign change of the MR at the pressure where the structural transition happens. The MR in the high-pressure range is positive and stable (Fig. 4b).

**Pressure and field dependences of the Hall coefficient**. Figure 4c shows magnetic field dependence of the Hall resistance in the pressure range of 6.1 − 30.0 GPa. At pressures below 10 GPa, clear kinks appear on the Hall resistance $R_{xy}$. The Hall coefficients $R_H$ are fitted in the low field (LF) range of 0 − 0.3 T and high field (HF) range of 0.5 − 1.0 T, respectively. The sign change of the $R_{HS}$ from positive in LF to negative in HF for the pressures at 6.1, 7.0, and 8.0 GPa demonstrates that the majority carriers change from holes to electrons as increasing magnetic field, indicating a Lifzshift of the Fermi surfaces induced by magnetic field. For high pressures (P > 10 GPa), the negative $R_{HS}$ for the LF and HF merge together gradually, revealing the majority carriers are electrons in the normal state of the SC phase. The densities of electrons estimated from the Hall coefficients change from ~$10^{19}$ to $10^{22}$ cm$^{-3}$ under pressure.

**Discussion**

$Cr_2Si_2Te_6$ and $FePSe_3$ have the same honeycomb lattice formed by the edge sharing octahedral centered by 3$d$ metal ions $Cr^{3+}$ or $Fe^{2+}$, respectively. $Cr_2Si_2Te_6$ is a FM insulator, where the electrical configuration of $Cr^{3+}$ ions is governed by the Hund's rule coupling[11,38]. The SC emerging coincides with the structural transition. Both the MR and Hall coefficient have clear changes as the phase transition. In contrast, $FePSe_3$ is an AFM insulator. Under pressure, SC appears concomitantly with a spin-crossover. Although mechanisms of the SC in AFM $FePSe_3$, FM $Cr_2Si_2Te_6$, and nonmagnetic $WTe_2$ are unclear, the superconducting state is in the vicinity of the ground states for the 2D vdW materials. Keeping the spin configuration $S = 0$ for the SC phase of $FePSe_3$ in mind, the $Fe^{2+}$ ions are nonmagnetic and differ from that of the iron-based superconductors. Although the spin configuration of $Cr^{3+}$ in the SC state of $Cr_2Si_2Te_6$ need to depict, the similarities of the SC in various in-plane honeycomb structures and vdW coupled lattices suggest a close relationship between the SC and structure.

In addition, the discovery of SC in the Cr-based honeycomb lattice $Cr_2Si_2Te_6$ under pressure, together with the SC in CrAs under pressure[40] and one-dimensional $K_2Cr_3As_3$ at ambient pressure[41], establishes a broad of Cr-based superconductors that could be used to explore SC and exotic physics.

**Methods**
**Material synthesis and characterization.** Bulk single crystals were grown by the flux method. The samples have been characterized previously. The single crystal XRD data at ambient pressure were taken at a single crystal x-ray diffractometer with the Cu $K\alpha$ ($\lambda$ = 0.1542 nm) radiation. The magnetization was measured by a Quantum Design physical property measurement system (PPMS).

**High-pressure structural characterization**. *In situ* high-pressure XRD patterns were collected at room temperature on the 4W2 beam line at Beijing Synchrotron Radiation Facility. The X-ray has the energy of 20 keV ($\lambda$=0.6199Å) and the diffraction patterns were recorded by using a PILATUS detector with exposure times of about 300 s. Powder samples were ground from single crystals at room temperature. A symmetric DAC with a pair of 400 μm diameter culet-sized diamond anvils was used for the *in situ* high pressure measurements. A steel gasket was pre-indented to 40 μm in thickness and the corresponding sample chamber with 150 μm in diameter was drilled by laser. A pre-compressed $Cr_2Si_2Te_6$ pellet was loaded in the middle of the sample chamber. Silicone oil was used as the pressure transmitting medium and the pressure in the diamond anvil cell was determined by measuring the shift of the fluorescence wavelength of the ruby spheres that were placed inside the sample chamber. The data were initially processed by using the Fit2d software[43] (with a $CeO_2$ calibration) and the subsequent refinements were calculated with Le Bail method using GSAS software[44,45].

**High-pressure transport measurements.** *In situ* high-pressure electrical resistance measurements on $Cr_2Si_2Te_6$ single crystals were carried out using a miniature DAC made from Be-Cu alloy and a PPMS. Diamond anvils with 400 μm culet were used, and the corresponding sample chamber with 150 μm in diameter were made in the insulating gasket. The gasket was achieved by a thin layered of cubic boron nitride and epoxy mixture. NaCl powders were employed as a pressure transmitting medium. Pressure was calibrated by using the ruby fluorescence shift at room temperature[46]. The standard four-probe technique was adopted in these measurements. The $Cr_2Si_2Te_6$ single crystals were cleaved along the *ab* plane with a typical size of 100×100×20 μm$^3$.

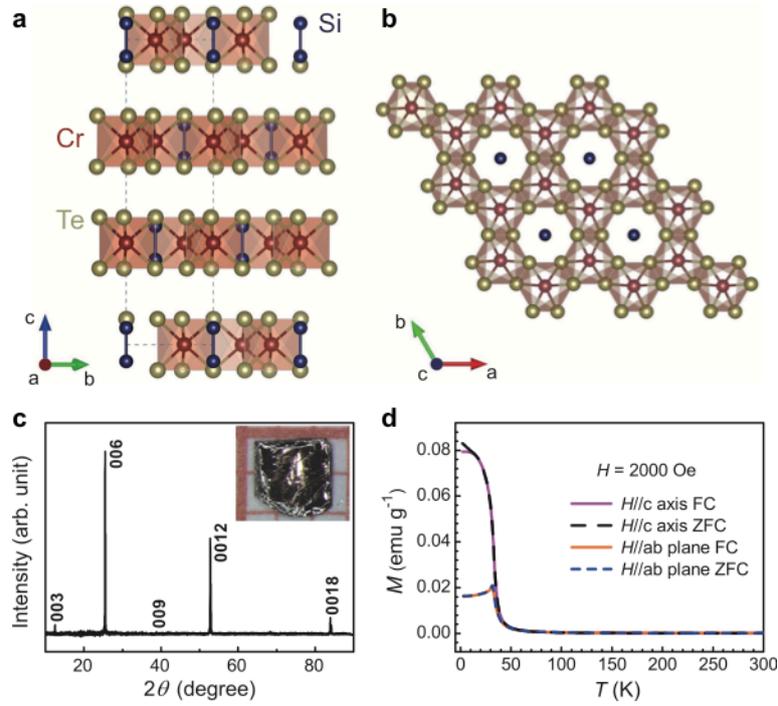

Figure 1 | Properties of $Cr_2Si_2Te_6$ at ambient pressure. Schematics of the crystal structure of $Cr_2Si_2Te_6$ viewed along (a) the *a* axis and (b) *c* axis, respectively. The dashed line in (a) indicates a unit cell. The view in (b) shows the honeycomb lattice. (c) Experimental and indexed XRD pattern measured on an exfoliated surface of a single crystal, as shown in the inset image. (d) Temperature dependence of the magnetization of $Cr_2Si_2Te_6$ on a single crystal under field cooling (FC) and zero field cooling (ZFC) modes in a magnetic field of 2000 Oe applied along the *c* axis and *ab* plane, respectively.

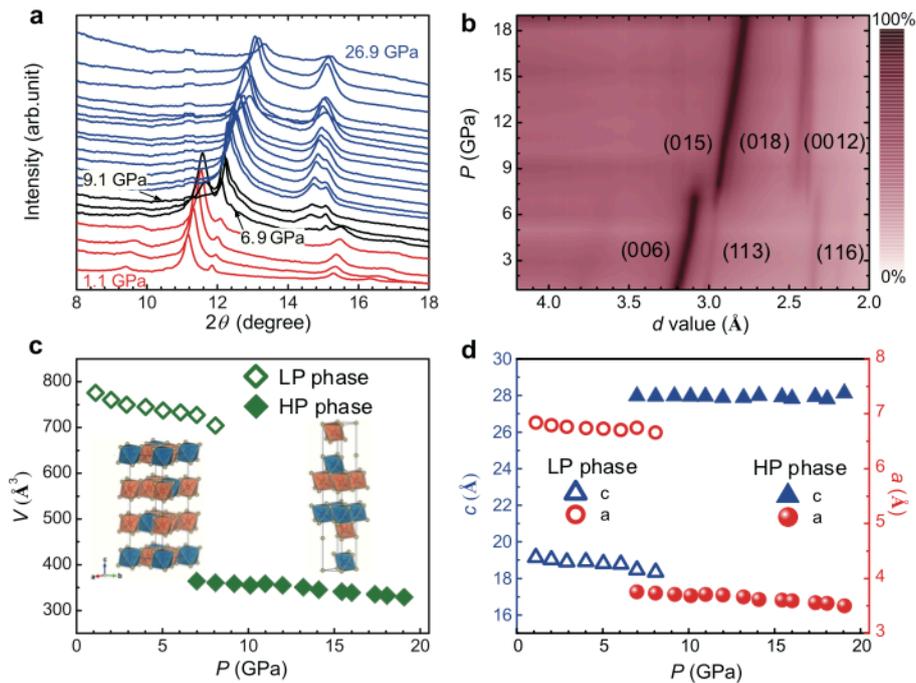

Figure 2 | Diffraction patterns and lattice constants of $Cr_2Si_2Te_6$ under pressure. (a) The XRD diffraction

patterns under pressures from 1.1 GPa to 26.9 GPa. A clear structural transition takes place at ~7 GPa. As increasing pressure, the diffraction peaks become weaker and broadened. (b) Two-dimensional XRD data showing abrupt changes for the (006), (113), and (116) peaks of $Cr_2Si_2Te_6$. The wavelength of the x rays is λ = 0.6199 Å. (c) The derived cell volume values as a function of applied pressure for the low pressure (LP) and high pressure (HP) phases of $Cr_2Si_2Te_6$. Insets show schematic views of the crystal structures at the LP and HP with the red, blue, and yellow globes representing the Cr, Si, and Te atoms, respectively. (d) Cell parameters of the LP and HP phases as a function of pressure.

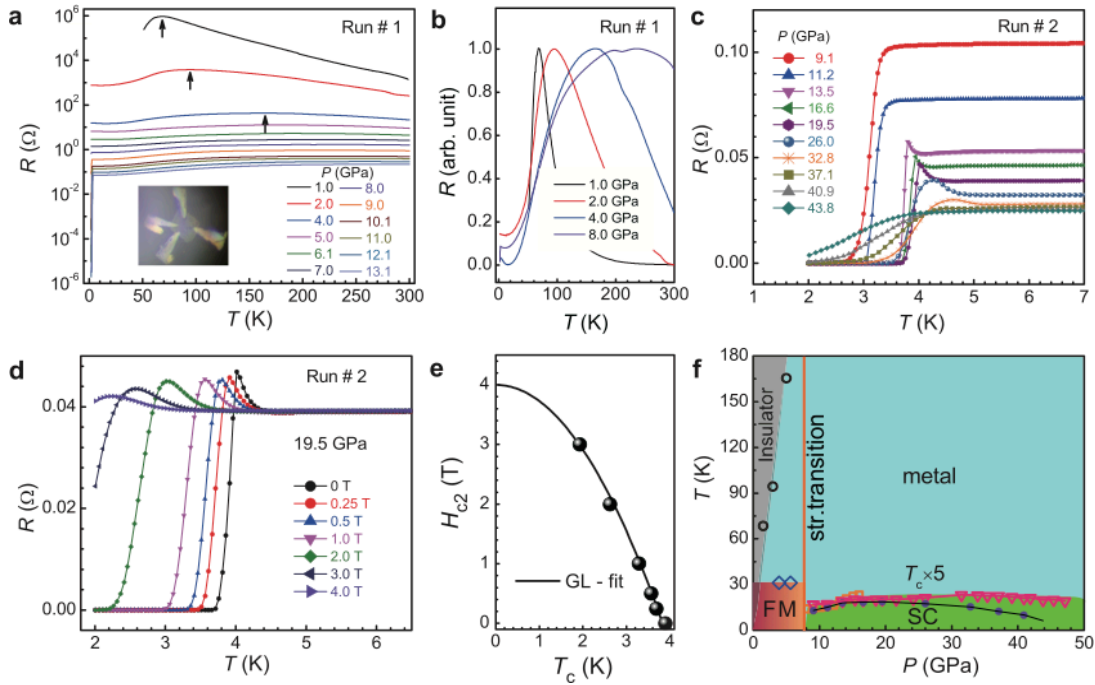

Figure 3 | Pressure dependence of the electrical resistance and a *T-P* phase diagram of $Cr_2Si_2Te_6$. (a) Temperature dependence of the resistance at pressures from 1.0 to 13.1 GPa shown on a logarithm scale. The arrows mark the humps in resistance. The inset is a photo of the sample and electrodes in the DAC. (b) Selected resistances at 1.0, 2.0, 4.0, and 8.0 GPa on a linear scale, showing the evolution of the IMT as functions of pressure and temperature. Superconductivity emerges at 8.0 GPa. (c) A zoom of the temperature dependence of resistances at pressures from 9.1 to 47.1 GPa, demonstrating SC occurs at ~3 − 4 K. (d) Superconducting transitions at 19.5 GPa under various magnetic fields. (e) The upper critical field $H_{c2}$ as a function of temperature at 19.5 GPa. The solid line represents a Ginzburg-Landau (GL) fitting. (f) A temperature-pressure phase diagram. The IMT and superconducting transition temperatures were determined from the resistance measurements on two samples. The structural (str.) transition pressure was measured with XRD at room temperature. The FM area is plotted based on the magnetic susceptibility at ambient pressure together with an indication of the Hall resistance measurements (See Supplementary). The $T_c$s have been multiplied by a factor of five. The purple points in the SC area are the $T_c$s determined from the temperatures with zero resistance.

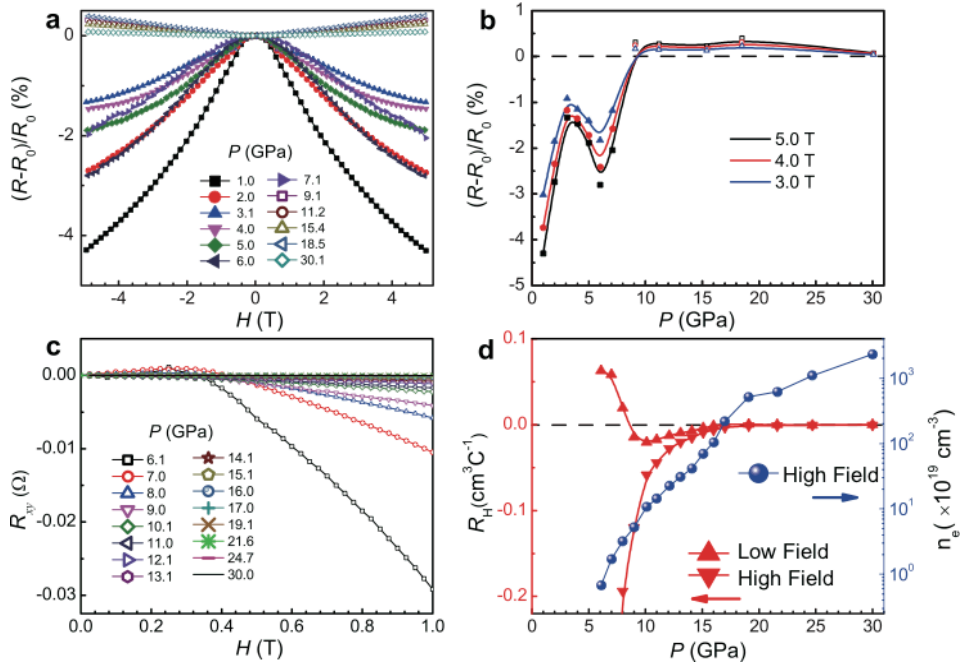

**Figure 4 | High-pressure magnetoresistance and Hall coefficient.** (a) Magnetic field dependence of the resistance changes $(R - R_0)/R_0$ measured at 10 K and different applied pressures, where $R$ is resistance and $R_0$ is the resistance at zero field. (b) Resistance changes $(R - R_0)/R_0$ as a function of pressure at 3.0, 4.0, and 5.0 T. The data were taken from (a) for two samples. The magnetoresistance changes sign at 10 GPa. The open points in (a) and (b) are from the first run and the solid points are from the third run. (c) Magnetic field dependence of the Hall resistance at different pressures and 10 K for the second run. (d) Hall coefficients as a function of pressure determined from the Hall resistance measured in (c). The low field data were fitted in the range of $0 < H \leq 0.3$ T, and the high field data were extracted from the range of $0.5 < H \leq 1.0$ T. The densities of electrons were obtained from the HF Hall coefficients.

**Acknowledgements**

The research was supported by the NSFC-11904414, NSFC-11904416, NSF of Guangdong under Contract No. 2018A030313055, the Fundamental Research Funds for the Central Universities (No. 18lgpy73), the Young Zhujiang Scholar program, and the Hundreds of Talents program of Sun Yat-Sen University. Y. G. acknowledges the supports from NSF of Shanghai (Grant No. 17ZR1443300) and NSFC-11874264. D. Y. is supported by NKRDPC-2018YFA0306001, NKRDPC-2017YFA0206203, NSFC-11574404, National Supercomputer Center in Guangzhou, and Leading Talent Program of

Correspondence and requests for materials should be addressed to M. W. (wangmeng5@mail.sysu.edu.cn) or to H. S. (sunhlei@mail.sysu.edu.cn).

# Supplementary: Pressure-induced superconductivity and structural transition in ferromagnetic Cr$_2$Si$_2$Te$_6$

Wanping Cai[1], Hualei Sun[1,*], Wei Xia[2], Changwei Wu[1], Ying Liu[3], Jia Yu[1], Junjie Yin[1], Yu Gong[4], Dao-Xin Yao[1], Yanfeng Guo[2], and Meng Wang[1,#]

[1]*School of Physics, Sun Yat-Sen University, Guangzhou, Guangdong 510275, China*
[2]*School of Physical Science and Technology, ShanghaiTech University, Shanghai 201210, China*
[3]*Xi'an Mordern Chemistry Research Institute, Xi'an, 710065, China*
[4]*Beijing Synchrotron Radiation Facility, Institute of High Energy Physics, Chinese Academy of Sciences, Beijing 100049, China*

During measurements of the Hall resistance (HR), we found a hysteresis of the HR for applying the magnetic field in two scanned directions. The HRs at different pressures for the second run were plotted in SFig. 1. The difference between the HRs in the negative magnetic field side deceases as increasing pressure and disappears at 9.0 GPa (SFig. 1d). The hysteresis may be related to the FM order in Cr$_2$Si$_2$Te$_6$.

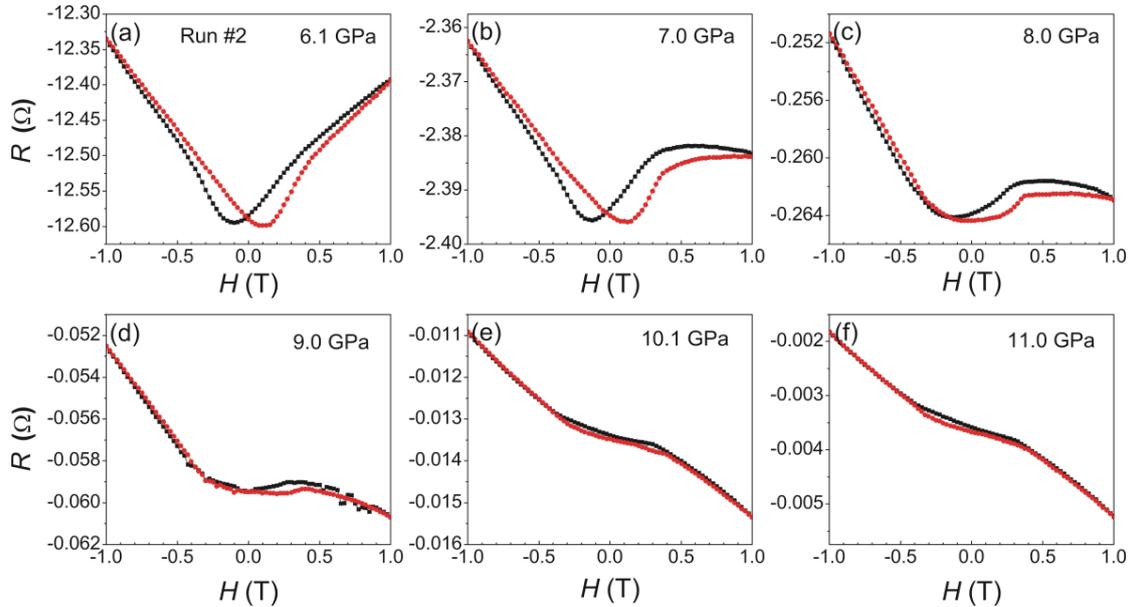

**SFigure 1 | Hall resistances measured at 10 K and various pressures for the second run.** (a) Hall resistances measured at 10 K and 6.1, (b) 7.0, (c) 8.0, (d) 9.0, (e) 10.1, and (f) 11.0 GPa. The black data points were collected for scanning the magnetic field from positive to negative. The red data points were measured reversely.

To confirm the hysteresis in HR, we did identical measurements in a new sample, named the fourth run, as shown in SFig. 2. The difference of the HR on the negative magnetic field side is clear at 10 K. As increasing temperature, the differences between the HR decrease and almost keep constant at 35, 40, and 45 K. The temperature dependence of the differences at 4.0 GPa suggests that the hysteresis is coupled to the FM order, and the FM ordered temperature is at ~33 K.

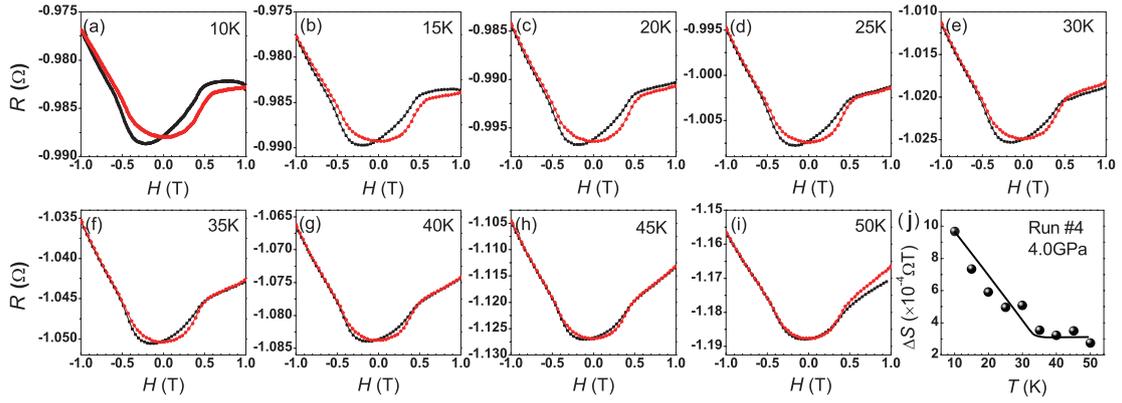

**SFigure 2 | Hall resistances measured at 4.0 GPa and various temperatures.** (a-i) Hall resistance measurements at 10, 15, 20, 25, 30, 35, 40, 45, and 50 K. The red and black data points were collected for scanning the magnetic field in two directions. (g) Differences of the integrated areas for two measured methods on the negative magnetic field side at the temperatures in (a-i). The solid line is a guide to the eyes. There is a kink at 32 K that is the FM ordered temperature at ambient pressure. The data were collected for the fourth run.

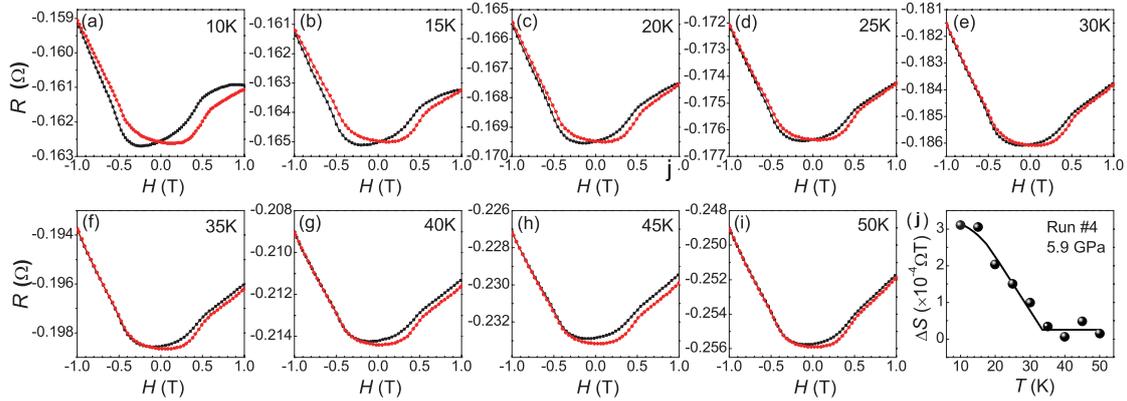

**SFigure 3 | Hall resistances measured at 5.9 GPa.** (a-g) Identical measurements and analysis for the fourth run as SFig. 2. The kink at ~32 K in (g) keeps at the same temperature as 4.0 GPa.

In SFig. 3, the pressure was fixed at 5.9 GPa. The hysteresis persists at 10 K and becomes absent at 35 K. Detailed measurements at various temperatures reveal a transition at ~33 K, indicating that the FM ordered temperature is not affected by the 5.9 GPa pressure. As the pressure increased to 9.9 GPa in SFig. 4, which is above the structural transition pressure of ~7.0 GPa, the hysteresis of resistances on the negative magnetic field side disappears. This is consistent with a nonmagnetic state after the structural transition and SC emerging.

We plot temperature dependence of the hysteresis for HR at 4.0, 5.9, and 9.9 GPa in SFig. 5. The results clearly demonstrate that the FM order persists up to 5.9 GPa, indicating that the FM transition as a function of pressure is first order and closely related to the structural transition and appearance of SC.

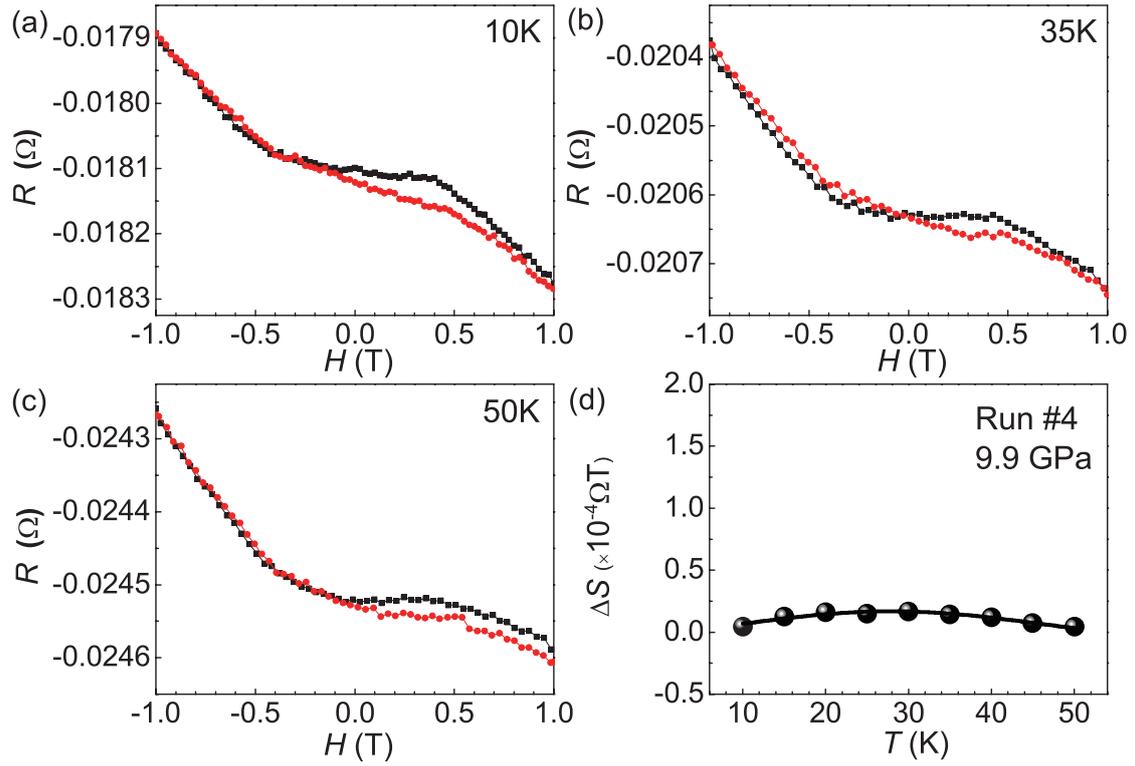

**SFigure 4 | Hall resistances measured at 9.9 GPa.** (a) Selected Hall resistances at 10 K, (b) 35 K, and (c) 50 K are presented. (d) Differences for the resistances measured with two scanned magnetic field directions in the negative magnetic field side as a function of temperature.

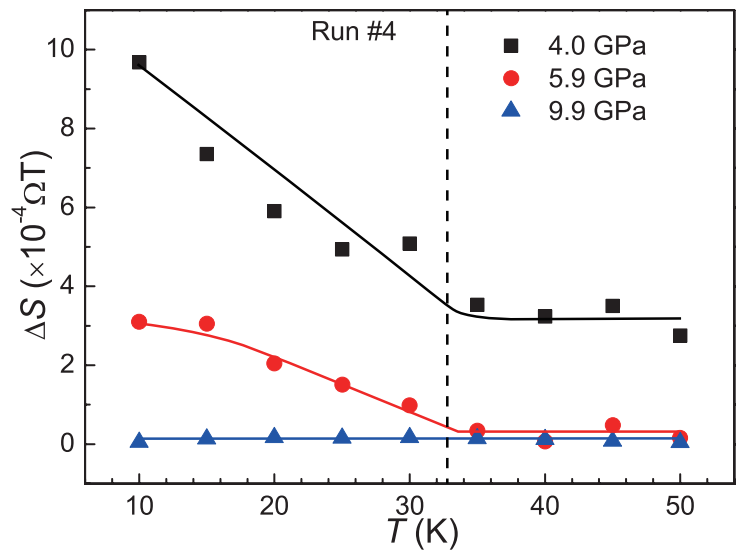

**SFigure 5 | A comparison of the Hall resistance hysteresis as a function of temperature at 4.0, 5.9, and 9.9 GPa for the fourth run.** The vertical dashed line denotes $T_N \sim 33$ K.